\documentclass[twocolumn,amsmath,amssymb,floatfix,prb]{revtex4}

\usepackage{graphicx}

\begin{document}

\title{Quantum dots investigated with charge detection techniques}

\author{T. Ihn, Simon Gustavsson,
Urszula Gasser, Bruno K\"ung,
Thomas M\"uller, Roland Schleser, Martin Sigrist, Ivan Shorubalko,
Renaud Leturcq, Klaus Ensslin}

\address{Solid State Physics Laboratory, ETH Zurich, CH-8093 Zurich}

\date{\today}

\begin{abstract}
The detection of the quantum dot charge state using a quantum point contact charge detector has opened a new exciting route for the investigation of quantum dot devices in recent years. In particular, time-resolved charge detection allowed the precise measurement of quantum dot shot noise at sub-femtoampere current levels, and the full counting statistics of the current. The technique can be applied to different material systems and holds promise for the future application in quantum dot based quantum information processing implementations. We review recent experiments employing this charge detection technique, including the self-interference of individual electrons and back-action phenomena.
\end{abstract}
\maketitle
\section{Introduction}
Quantum dots are semiconductor nanostructures of sizes ranging between a few to a few hundred nanometers in which electrons, holes, or excitons are confined so strongly in all three spatial directions that the charging energy and the quantum confinement energy are of similar magnitude. The term quantum dot was coined \cite{Reed86} as an extrapolation from one-dimensional confinement in quantum wells, via two-dimensional confinement in quantum wires to complete three-dimensional confinement in quantum dots. This happened at a time when nanofabrication technology made such rapid progress that the control over semiconductor nanostructure devices was tremendously improved enabling the measurement of the Coulomb blockade effect in electron transport experiments at liquid Helium temperatures and below. Since then the charge states \cite{Kouwenhoven97}, and recently also the spin states of quantum dots \cite{Hanson07} were heavily investigated experimentally. The suggestion to use semiconductor quantum dots as spin-qubits \cite{Loss98} lead to the development of well-controlled few-electron quantum dots \cite{Kouwenhoven01,Ciorga00,Sprinzak02,Sigrist06}, also called artificial atoms.

\section{Time averaged charge detection}
\label{}
\subsection{Overview}
Monitoring quantum dot charging with individual electrons using a quantum point contact charge detector
was introduced in 1993 by Field and coworkers \cite{Field93}. This early experiment was designed to detect the oscillatory electrostatic potential variations of a quantum dot tuned through Coulomb blockade oscillations.
The technique was later used in experimental designs aiming at controlled dephasing of electrons in an interferometer \cite{Buks98}, and at the measurement of the charge distribution in a Kondo-correlated few-electron quantum dot \cite{Sprinzak02}. In the latter experiment it was demonstrated that a charge detector allows to determine the absolute number of electrons in a few-electron quantum dot, even if the direct current through the dot is too small to be measured. The application of the technique was then extended to double quantum dot devices \cite{Smith02,Elzerman03,Petta04}, where it can serve to establish the regime, where each of the two quantum dots is occupied only with a single electron \cite{Elzerman03,Petta04}. At that time, also the theory of quantum measurement with quantum point contact detectors was investigated \cite{Aleiner97,Levinson97,Gurvitz97,Buttiker00,Silva01,Korotkov01,Goan01,Clerk03,Buttiker03}.

\subsection{Principle of operation}

\begin{figure}[tbp]
\includegraphics[width=8cm]{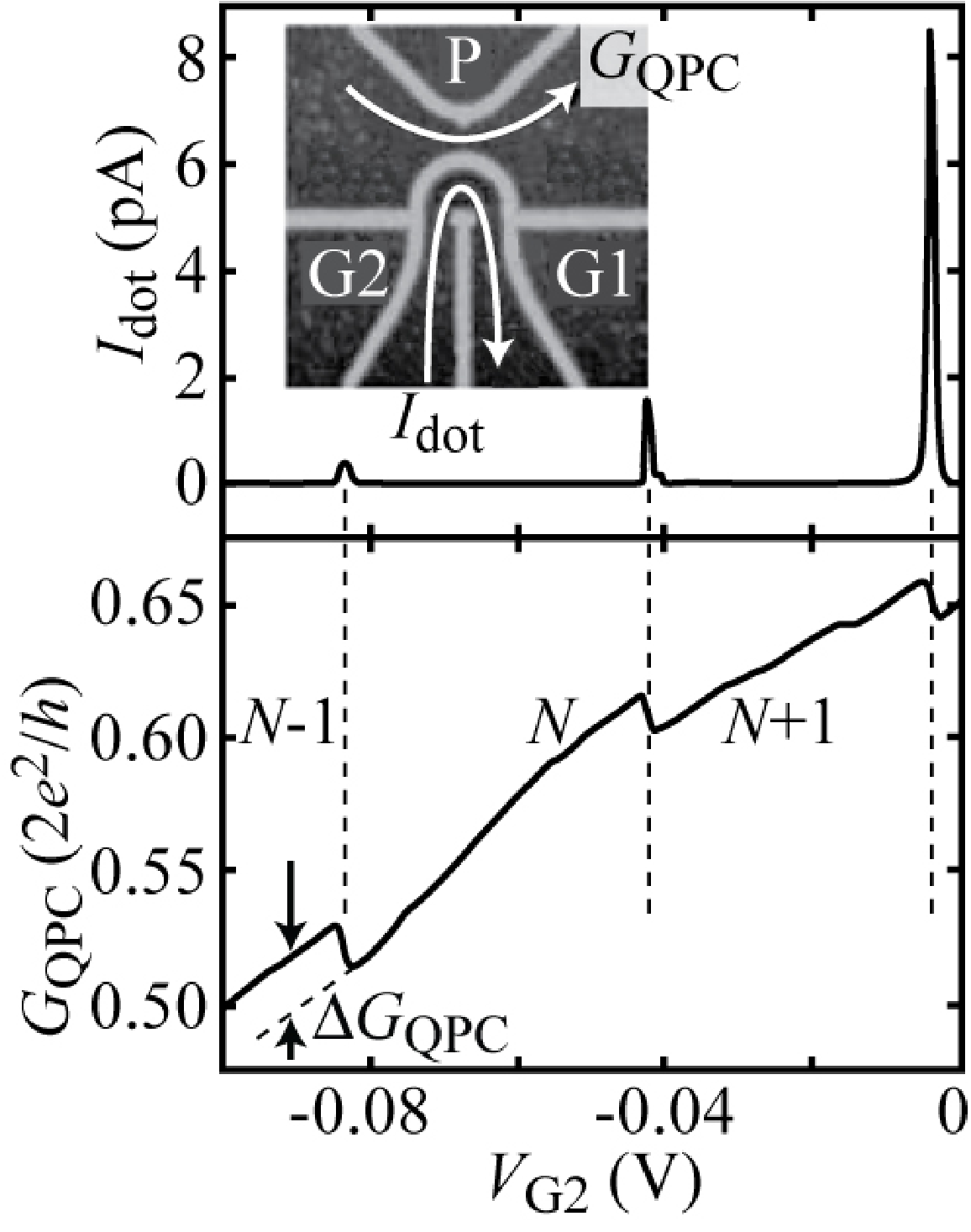}
\caption{Operation principle of a quantum dot charge detector. The inset shows the quantum dot sample tunable with gates G1 and G2. The charge sensor is capacitively coupled to the dot, and tunable with gate P. The upper panel shows the current $I_\mathrm{dot}$ through the quantum dot exhibiting three conductance resonances as a function of the voltage $V_\mathrm{G2}$ applied to G2. The lower panel shows the conductance $G_\mathrm{QPC}$ of the quantum point contact detecting the addition of individual electrons to the quantum dot.}
\label{fig1}
\end{figure}

The inset of Figure\,\ref{fig1} shows a scanning force microscope image of a sample fabricated on the basis of a Ga[Al]As heterostructure containing a two-dimensional electron gas 34\,nm below the surface \cite{Schleser04}. The surface has been patterned by local anodic oxidation to form a quantum dot structure, an adjacent quantum point contact, and in-plane gates P, G1, and G2. The oxide lines, appearing bright in the image, deplete the electron gas below. The structure allows to drive a current through two separate electronic circuits as indicated by the white arrows. On one hand, the quantum dot current $I_\mathrm{dot}$ is measured. It exhibits Coulomb blockade with intermittent resonances at gate voltages $V_\mathrm{G2}$ where the electron number in the dot is increased by one. On the other hand, the conductance $G_\mathrm{QPC}$ of the quantum point contact is measured. No direct current can flow from one circuit to the other. Whenever a negatively charged electron is added to the quantum dot, the conductance of the quantum point contact is reduced by $\Delta G_\mathrm{QPC}$ as a result of the repulsive Coulomb potential created in the quantum point contact (see dashed lines). The overall slope of $G_\mathrm{QPC}$ is the result of direct capacitive coupling between the quantum point contact and the gate G2. For a given geometry, the best sensitivity of the charge detector is achieved, if it is operated at the point of maximum slope in the $G_\mathrm{QPC}$ vs. $V_\mathrm{P}$ characteristic at a conductance between complete pinch-off and the first conductance plateau at $2e^2/h$ (not shown). The charge detector interacts most strongly with the quantum dot system, if it is situated as close as possible to the dot, and if metallic top gates, which screen the interaction, are either entirely avoided (as in the structure shown in the inset of Fig.\,\ref{fig1}), or at least minimized in area. The signal-to-noise in the measured quantum point contact current can be maximized by increasing the source--drain bias voltage, however, the upper limit is given by the magnitude of the quantization energy transverse to the current flow which is typically of the order of 1\,meV in Ga[Al]As devices. As a consequence, typical source--drain voltages are well below 1\,mV, and typical currents stay below 100\,nA. The noise in the detector circuit is usually governed by the noise of the low-frequency current--voltage converter which, connected to the highly capacitive wiring of a dilution refrigerator, can suffer from a considerable capacitive noise gain.

Measuring the quantum dot current alone does not allow to extract the strengths of the tunneling coupling of quantum dot states to source and drain contacts $\Gamma_\mathrm{S}$ and $\Gamma_\mathrm{D}$. In the single-level transport regime of the quantum dot, the additional measurement of the quantum point contact conductance gives additional information allowing to extract $\Gamma_\mathrm{S}$ and $\Gamma_\mathrm{D}$ for ground states, and in favorite cases even for excited states \cite{Schleser05,Rogge05}.

Quantum point contact charge detectors have  been employed for the measurement of charge rearrangements within a quantum dot at high magnetic fields and constant electron number, where spatially separate edge channels, so-called Landau shells, exist \cite{Fricke05}. Different Landau shells couple with different capacitance to the charge detector and can thereby be distinguished by the detector.

Naturally, a quantum point contact charge detector is also sensitive to undesirable charging of impurity sites in the vicinity of the quantum dot system of interest. Such charge rearrangements are known to spoil conventional conductance measurements of a quantum dot. Using a dot--quantum point contact arrangement in a scanning gate experiment, such impurity centers could be localized in real space, and their density could be estimated \cite{Gildemeister07}.

The principle of charge detection is not limited to the Ga[Al]As material system. The technique has, for example, been successfully applied to quantum dots in InAs nanowires \cite{Shorubalko08}, in Si/SiGe \cite{Simmons07},  and to quantum dots in graphene \cite{Guttinger08}. In the graphene experiment, the constriction used for charge detection does, however, not exhibit conductance quantization, but the strong potential fluctuations in the constriction rather lead to localization of charge carriers which manifests itself in conductance resonances measured as a function of the plunger gate voltage. The steep slopes of these resonances gives excellent charge sensitivity, similar to charge detection experiments with quantum dots \cite{Duncan99} or single-electron transistors \cite{Ilani01,Ilani04}.

\section{Time-resolved charge detection}

First time-resolved charge detection measurements on quantum dots were not performed with quantum point contact detectors, but with radio-frequency single-electron transistors \cite{Lu03,Fujisawa04,Bylander05}.
Time-resolved measurements with quantum point contact charge detectors started with tuning up the bandwidth of conventional low-frequency setups \cite{Schleser04}. As mentioned above, for a given cryostat wiring, the capacitive noise gain of the current--voltage amplifier limits the achievable bandwidth. Using such setups, bandwidths of up to 30--40\,kHz have been reported in the literature \cite{Elzerman04,Gustavsson06}, limiting the time resolution to the order of ten microseconds.

\begin{figure}[tbp]
\includegraphics[width=8cm]{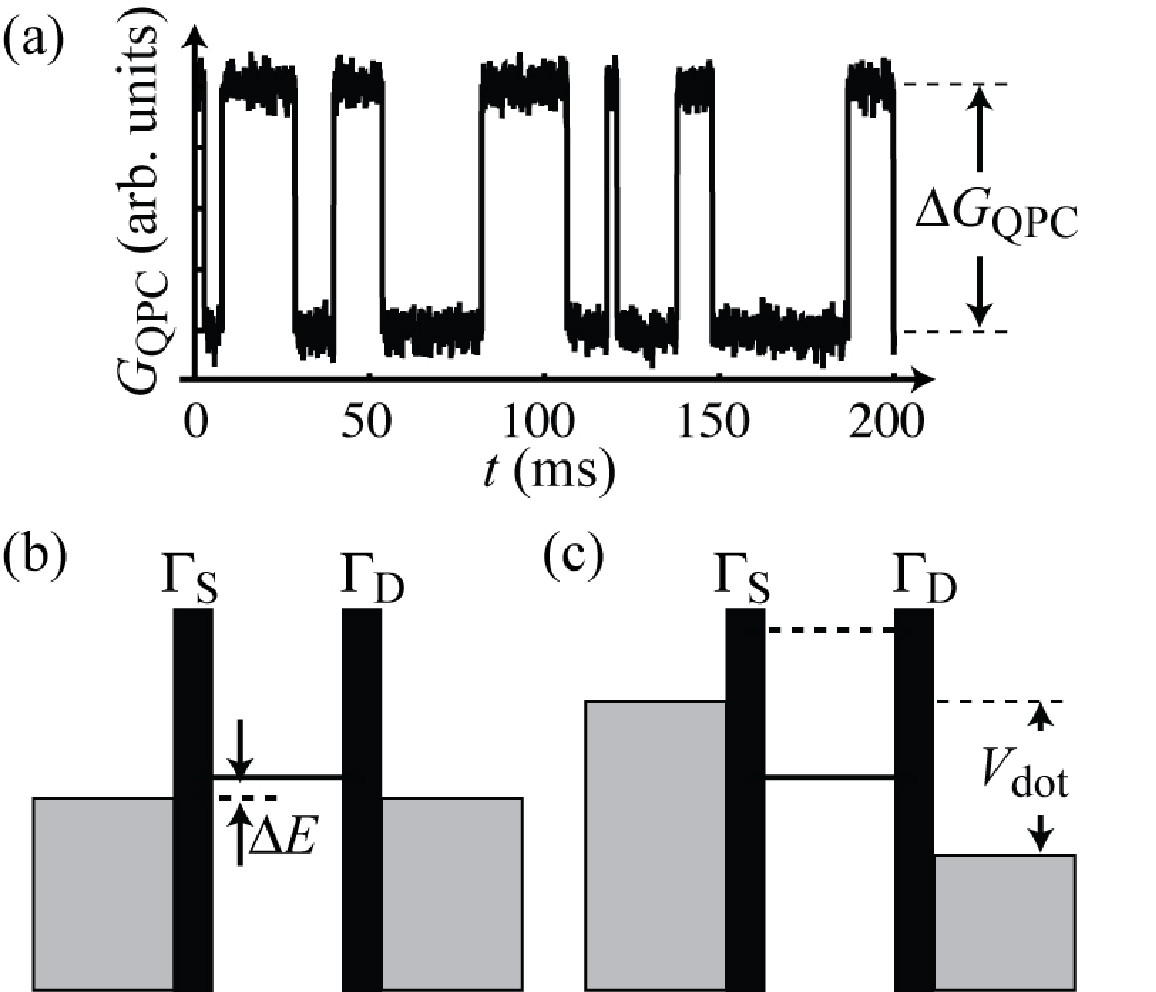}
\caption{(a) Time resolved signal of a quantum point contact detector measured close to a conductance resonance of the quantum dot. The signal switches randomly by $\Delta G_\mathrm{QPC}$ between two distinct levels. (b) Schematic energy diagram of the quantum dot showing the situation $V_\mathrm{dot}=0$. (c) The same for $V_\mathrm{dot}\gg k_\mathrm{B}T$.}
\label{fig2}
\end{figure}

In measurements with time-independent gate voltages, the charge detector witnesses electrons tunneling into and out of the quantum dot in real time. This manifests itself in random switching of the detector conductance between two distinct levels as shown in Fig.\,\ref{fig2}(a). When the conductance switches downwards, an electron has entered the dot, if it switches upwards, an electron has left the dot. In most cases the analysis assumes that the quantum dot is in the single-level transport regime. In this case, the time-separations between tunneling-in and tunneling-out events follow the exponential decay laws
\[ p_\mathrm{in/out}(t)dt = \Gamma_\mathrm{in/out}e^{-\Gamma_\mathrm{in/out}t}dt\]
with characteristic tunneling-in an tunneling-out rates $\Gamma_\mathrm{in/out}$. This decay law has been confirmed experimentally \cite{Schleser04,Gustavsson06,MacLean07}. The autocorrelation function of such a two-level random signal has been calculated by Machlup in 1954 \cite{Machlup54}.
The interpretation of the rates $\Gamma_\mathrm{in}$ and $\Gamma_\mathrm{out}$ depends on the source--drain bias voltage $V_\mathrm{dot}$ applied to the quantum dot.

If $V_\mathrm{dot}=0$ [see Fig.\,\ref{fig2}(b)], the charge detector delivers the signal of the equilibrium fluctuations (thermal noise) of charge carriers \cite{Schleser04,Gustavsson06a}. In this case, $\Gamma_\mathrm{in} = \Gamma f(\Delta E/k_\mathrm{B}T)$, and $\Gamma_\mathrm{out} = \Gamma[1- f(\Delta E/k_\mathrm{B}T)]$, with $\Gamma = \Gamma_\mathrm{S}+\Gamma_\mathrm{D}$ being the total dot--lead coupling, $f(x)=[\exp(x)+1]^{-1}$ the Fermi--Dirac distribution function, $\Delta E$ the difference between the electrochemical potential in the leads and that in the dot, and $T$ the electron temperature in the leads. A shortcoming of the charge detection technique applied in this way is that the detector is insensitive to whether the tunneling electron originated from the source or from the drain contact. Tunneling rates can be individually determined, if the respective other barrier is deliberately pinched off. In addition, this measurement delivers the electron temperature $T$ of the leads, if the lever arm of the gate is known from Coulomb blockade diamond measurements. It is straightforward to see how the analysis will change, if the involved dot level is spin degenerate \cite{Gustavsson06}.

In contrast, if $V_\mathrm{dot}\gg k_\mathrm{B}T$, but only a single quantum state is in the bias window [see Fig.\,\ref{fig2}(c)], the rates $\Gamma_\mathrm{in/out}$ obtained from a time trace can be interpreted directly as the tunneling rates $\Gamma_\mathrm{S}$ and $\Gamma_\mathrm{D}$ \cite{Gustavsson06,Gustavsson06a}. In this case, the electron tunneling into the dot will always originate from the source contact, and it will always tunnel out to the drain. A detailed analysis of the energy dependence of tunneling rates was performed by MacLean and coworkers \cite{MacLean07}. The case of more than a single energy level in the bias window was investigated by Gustavsson and coworkers in Ref.\,\cite{Gustavsson06a}.

\begin{figure}[tbp]
\includegraphics[width=8cm]{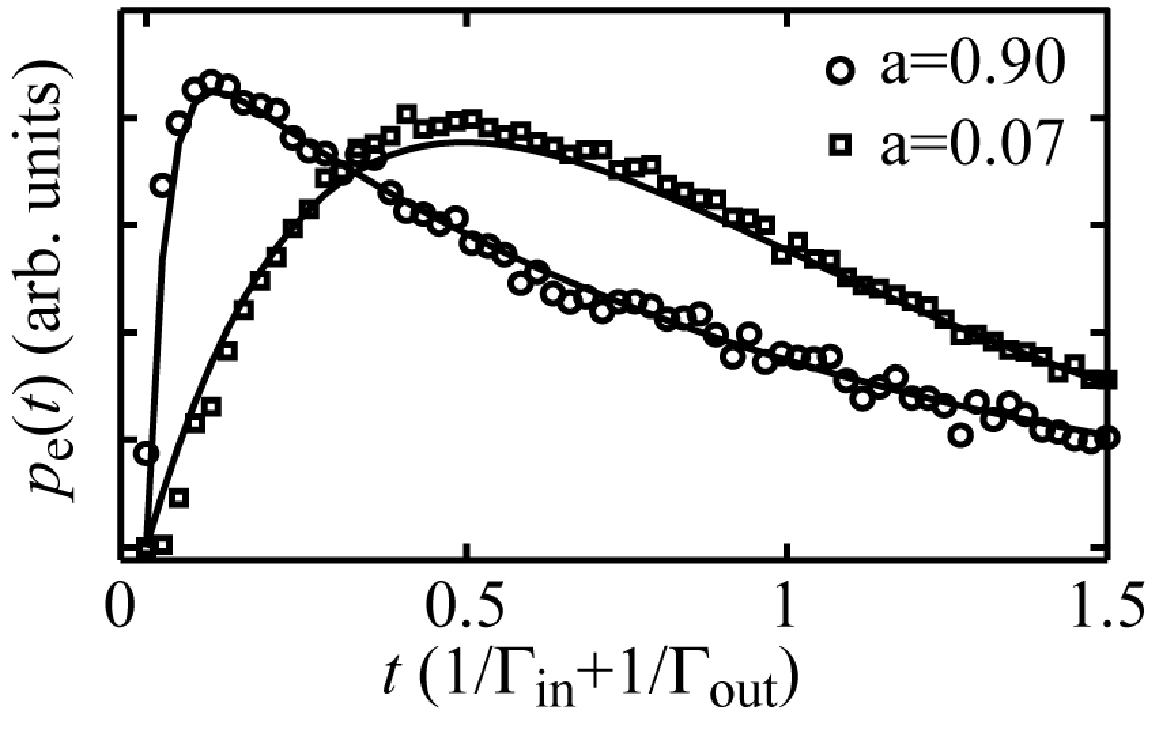}
\caption{Distribution $p_\mathrm{e}(t)$ of times needed for one electron to traverse the quantum dot. Symbols are measured data points, solid lines are predictions of theory. The two distributions corresponding to different coupling asymmetries $a$ (see text) are plotted on different vertical scales for clarity.}
\label{fig3}
\end{figure}

On the next level, correlations between subsequent tunneling-in and tunneling-out events at $V_\mathrm{dot}\gg k_\mathrm{B}T$ can be considered. For example, if we assume that such pairs of subsequent in/out events (in the following we call the pair {\em event} for simplicity) are statistically independent, we find the statistical distribution
\begin{multline*}
p_\mathrm{e}(t)dt = dt\int_0^t dt' p_\mathrm{in}(t')p_\mathrm{out}(t-t') \\
= \frac{\Gamma_\mathrm{in}\Gamma_\mathrm{out}}{\Gamma_\mathrm{in}-\Gamma_\mathrm{out}}\left(e^{-\Gamma_\mathrm{in}t}-e^{-\Gamma_\mathrm{out}t}\right)dt.
\end{multline*}
Figure\,\ref{fig3} shows measurements of this distribution function for two different coupling asymmetries $a=(\Gamma_\mathrm{in}-\Gamma_\mathrm{out})/(\Gamma_\mathrm{in}+\Gamma_\mathrm{out})$. For almost symmetric coupling ($a=0.07$) of the dot to the source and drain lead, there is a pronounced suppression of the distribution for small times. This is a direct consequence of the correlation between subsequently tunneling electrons brought about by the Coulomb blockade effect. The second electron has to wait with tunneling in until the first electron has tunneled out of the dot. This suppression becomes narrower in time for strongly asymmetric coupling ($a=0.90$), because the system approaches the limit of a single barrier device in which no Coulomb blockade exists.

Counting electrons traversing quantum dots under the condition $V_\mathrm{dot}\gg k_\mathrm{B}T$ is a direct way to measure the electrical current. While the method finds its upper bound at the few femtoampere level as a result of the finite bandwidth of the detector circuit, the lower bound of measurable currents is essentially given by the stability of the sample (which can be more than months for a good one), and by the patience of the observer. The acuracy of the current measurement scales with the number of counts $N$ as $1/\sqrt{N}$. As a result, attoampere current levels can be measured with a one percent accuracy within one hour, 0.2\,aA within one day.

A different class of time-resolved charge detection experiments that we will not treat in detail here, are pulsed gate measurements. The time-resolution of the quantum point contact charge detector allows to realize a so-called single-shot charge readout of a quantum dot qubit \cite{Elzerman04}. Using this technique, the spin of an electron injected into a quantum dot can be measured (spin-to-charge conversion technique). In addition these experiments give access to the spin relaxation time $T_1$ in quantum dots. The relaxation times between Zeeman-split levels of a one-electron quantum dot were measured to be of the order of 1\,ms at a magnetic field of 8\,T \cite{Elzerman04}. In the most recent experiments \cite{Amasha08} it was found that $T_1$ is dominated by spin--orbit mediated coupling to phonons, a mechanism that can be modified by gate voltages that influence the orbital confinement of the electron. The values of $T_1$ could be changed by more than one order of magnitude with maximum values of more than 1\,s. In Ref.\,\cite{Hanson05} singlet--triplet relaxation times of several milliseconds were found in a two-electron quantum dot using the same technique. A theoretical discussion of the single-shot readout method is given in Ref.\,\cite{Engel04}. 

Time-resolved charge detection techniques have also been applied to other material systems, such as quantum dots in InAs nanowires \cite{Gustavsson08d}, and they have been extended to double quantum dot systems \cite{Fujisawa06,Gustavsson08,Choi09}. Double quantum dots are of particular interest for several reasons that will become more transparent below. At this point we highlight the fact that they allow bidirectional electron counting, i.e., the direction of electron tunneling can be determined on the single electron level \cite{Fujisawa06}. Using two quantum point contact charge detectors coupled to a double quantum dot, cross correlation techniques can be applied to significantly improve the signal-to-noise ratio for charge detection \cite{Kung09}. This allows to reduce the invasiveness of the detection process, and may help to increase the available detection bandwidth in setups that are noise-limited.

Time-resolved charge detection can be made considerably faster by using amplifier setups with larger bandwidths. The low-frequency setups described above suffer from the unavoidable cable capacitances at the input of the current--voltage amplifier. Significant bandwidth increase to 1\,MHz has been achieved with a cryogenic preamplifier \cite{Lee89} operating at a temperature of 1\,K \cite{Vink07}. This allows mounting the amplifier closer to the sample, thus reducing the capacitive load, and the low temperature reduces the amplifier noise. The dissipated amplifier power of 30\,$\mu$W can easily be cooled away. However, an even stronger increase in measurement bandwidth has been demonstrated with radio-frequency (RF) quantum point contact setups \cite{Muller07,Reilly07,Cassidy07,Thalakulam07}. Within this approach, the quantum point contact impedance, which is about 25\,k$\Omega$ at the operating point, is matched to the 50\,$\Omega$ impedance of the coaxial cables with a suitable $L-C$ matching circuit. Resistance changes of the quantum point contact are seen as changes in the reflected RF-power. The reflected signal is split from the incoming wave by a directional coupler, and can then be amplified with a commercial cryogenic RF-amplifier. M\"uller and coworkers reported a charge detection bandwidth of 10\,MHz corresponding to a time resolution of 50\,ns limited by their demodulation setup at a carrier frequency of about 200\,MHz. Later, twice this value was reported in Ref.\,\cite{Cassidy07} for a carrier frequency around 300\,MHz. In the ultimate limit, the noise performance of the quantum point contact charge detector is limited by its own shot noise, as demonstrated in Ref.\,\cite{Thalakulam07}. In the following we estimate the maximum bandwidth for a shot noise limited quantum point contact charge detector: the shot noise itself is given by $\Delta I_\mathrm{n}=\sqrt{2eI_\mathrm{QPC}\Delta f}$. The amplitude of the switching signal is estimated to be $\Delta I_\mathrm{S}\approx s I_\mathrm{QPC}\approx s V_\mathrm{QPC}e^2/h$ with $s\approx 3\%$ being a reasonable estimate for the relative change of the quantum point contact conductance, and $V_\mathrm{QPC}\approx 1$\,mV. Counting is possible for $\Delta I_\mathrm{n}\ll \Delta I_\mathrm{S}$, giving a maximum bandwidth well below 100\,MHz. It seems therefore that Ref. \cite{Thalakulam07} was already reasonably close to the maximum bandwidth. Room for improvement is, of course, in the coupling strength between quantum dot and detector. In systems such as InAs nanowire quantum dots read out with a GaAs charge detector in an underlying two-dimensional electron gas, the $s$ can reach values well above 50\% \cite{Gustavsson08d}.

\section{Shot noise and full counting statistics}

\begin{figure}[tbp]
\includegraphics[width=8cm]{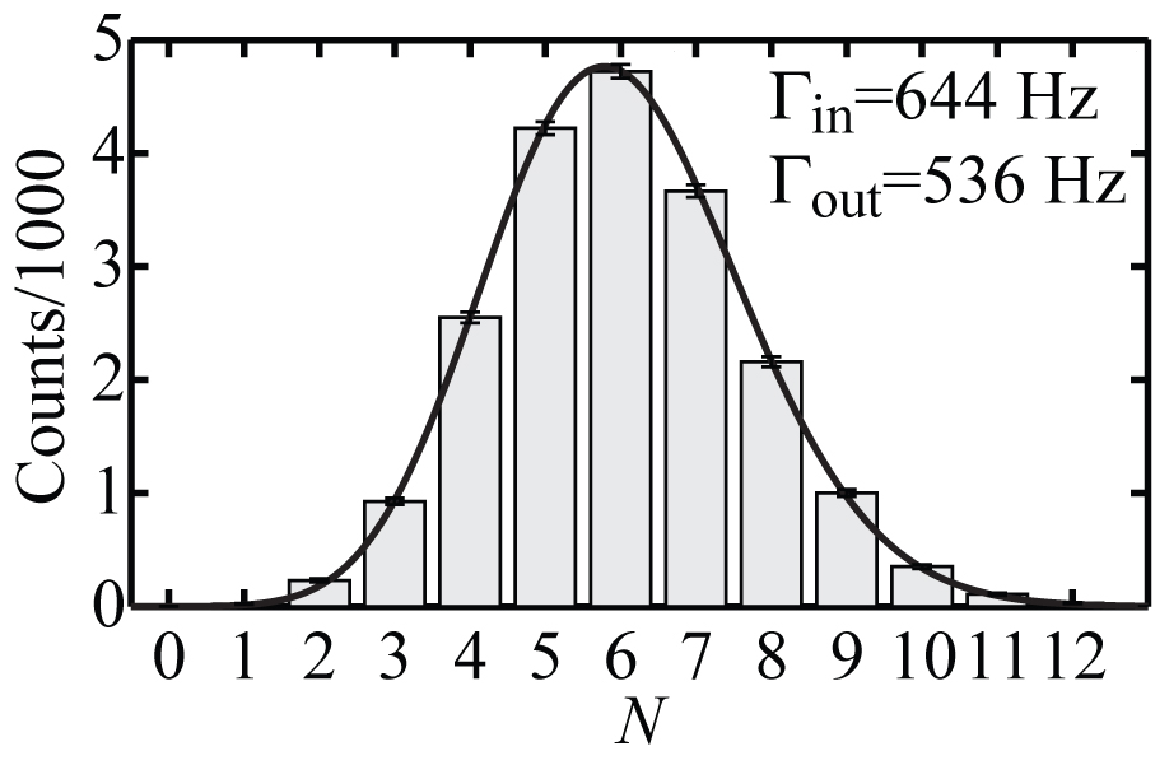}
\caption{Histogram of the full counting statistics of a quantum dot. The solid line is the theoretical prediction for the given rates $\Gamma_\mathrm{in}$ and $\Gamma_\mathrm{out}$.}
\label{fig4}
\end{figure}

An alternative way of analyzing time-resolved single-electron tunneling traces such as that shown in Fig.\,\ref{fig2}(a) is called full counting statistics. In order to do this analysis, a time trace of length $T$ is divided into a reasonably large number of shorter segments of equal length $\Delta T$. A histogram is then plotted for the distribution of the number $N$ of events found in the segments (an event is, for example, a down-switch of $G_\mathrm{QPC}$). An example of such a histogram, similar to those reported by Gustavsson in Ref.\,\cite{Gustavsson06} is shown in Fig.\,\ref{fig4}. The mean value (first moment) $\langle N\rangle$ calculated with this histogram gives the mean current $I_\mathrm{dot}=e\langle N\rangle/\Delta T$ through the quantum dot. However, the width of the histogram, characterized by its second central moment (variance) $\langle(N-\langle N\rangle)^2\rangle$ is a measure for the fluctuations $\langle\Delta I^2\rangle=e^2\langle(N-\langle N\rangle)^2\rangle/\Delta T$ of the quantum dot current, meaning its shot noise. The shot noise for quantum dots has been calculated in Ref. \cite{Davies92} and later discussed in the framework of full counting statistics \cite{Bagrets03}.
While the shot noise of a single barrier device is expected to follow poissonian statistics with $\langle N\rangle=\langle(N-\langle N\rangle)^2\rangle$ (the mean equals the variance), for quantum dots the shot noise is expected to be suppressed as a result of the Coulomb interaction-mediated correlations between tunneling electrons [see also the suppression of the distribution in Fig.\,\ref{fig3} at short times]. From Fig.\,\ref{fig4} a variance $\langle(N-\langle N\rangle)^2\rangle\approx 3$ can be estimated, compared to a mean $\langle N\rangle\approx 6$, implying a reduction of the width by a factor $F=1/2$ compared to the poissonian case. The quantity $F$ is called the Fano factor. Given the histogram shown in Fig.\,\ref{fig4}, even higher central moments, such as the skewness (3rd central moment) or the kurtosis (4th central moment) can be experimentally determined. In Ref.\,\cite{Gustavsson07a} it was possible to determine all cumulants up to the fifths reliably from the experiment. In order to achieve this accuracy, systematic corrections due to the finite bandwidth of the detector circuit had to be taken into account \cite{Naaman06}.

The full counting statistics can be found theoretically from a master equation approach. For example, in the single-level transport regime, the quantum dot system may be described by a two-state system with state 0 denoting zero, state 1 denoting one excess electron in the dot. We measure the current by counting the number $N$ of electrons that transmit through the dot--drain barrier. We consider the case $V_\mathrm{dot}\gg k_\mathrm{B}T$ as depicted in Fig.\,\ref{fig2}(c), such that tunneling-in is only possible from the source (rate $\Gamma_\mathrm{S}$), and tunneling-out only through the drain (rate $\Gamma_\mathrm{D}$). The master equation is then given by
\begin{eqnarray*}
dp_0(t|N)/dt & = & -\Gamma_\mathrm{S}p_0(t|N)+\Gamma_\mathrm{D}p_1(t|N-1)\\
dp_1(t|N)/dt & = & -\Gamma_\mathrm{D}p_1(t|N)+\Gamma_\mathrm{S}p_0(t|N)
\end{eqnarray*}
Here, $p_n(t|N)$ is the probability that at time $t$, the system is found in state $n$, given that $N$ electrons have been transferred into the drain lead since $t=0$. At $t=0$ we have the initial conditions $p_0(t=0|N=0)=1$, and $p_n(t=0|N\neq 0)=0$. The rate equation can be solved using the discrete Fourier transform $p_n(t|\chi) = \sum_N p_n(t|N)\exp(iN\chi)$, where $\chi$ is called the counting field. One finds the linear differential equation
\[
\frac{d}{dt}\left(\begin{array}{c}p_0(t|\chi)\\p_1(t|\chi)\end{array}\right) = 
\left(\begin{array}{cc} -\Gamma_\mathrm{S} & \Gamma_\mathrm{D}e^{i\chi}\\ \Gamma_\mathrm{S} & -\Gamma_\mathrm{D} \end{array}\right)\left(\begin{array}{c}p_0(t|\chi)\\p_1(t|\chi)\end{array}\right),
\]
which has the general solution $p_n(t|\chi) = \sum_{j=0}^1 c_{nj}\exp[\lambda_j(\chi) t]$ with the $\lambda_j(\chi)$ being the eigenvalues of the coefficient matrix. For times $t$ large compared to the correlation time $(\Gamma_\mathrm{S}+\Gamma_\mathrm{D})^{-1}$ \cite{Machlup54}, the solution is governed by the eigenvalue with the smallest negative real part (say, $\lambda_0$) giving the slowest decay. The full counting statistics, i.e., the probability that $N$ electrons have been transferred through the dot after time $\Delta T$ is given by
\[ P_N(\Delta T) = \sum_{n=0}^1 p_n(\Delta T|N)=\frac{1}{2\pi}\int d\chi e^{-iN\chi} \sum_{n=0}^{1}p_n(\Delta T|\chi). \]
The logarithm of its Fourier transform is the cumulant generating function $S(\chi)$, which has the large $\Delta T$ limit $ S_{\Delta T}(\chi) = \lambda_0(\chi) \Delta T$. 
The mean current is given by the first cumulant $\langle N\rangle = -idS/d\chi |_{\chi=0}$, the shot noise by the second cumulant $\langle(N-\langle N\rangle)^2\rangle = -d^2S/d\chi^2|_{\chi=0}$.
The resulting full counting statistics, which has been worked out by Bagrets and Nazarov \cite{Bagrets03}, is plotted as a solid line in Fig.\,\ref{fig4}, and shows excellent agreement with the measured histogram. Finite bandwidth corrections \cite{Naaman06} have been taken into account. More details about the analysis of full counting statistics data can be found in the review \cite{Gustavsson07b}, and in the overview article \cite{Gustavsson08e}. A theoretical discussion of the joint current probability distribution describing the connection between the acquisition of information by detection and the uncertainty in the system can be found in Ref.\,\cite{Sukhorukov07}.

\section{Self-interference of individual electrons detected by electron counting}

We continue by discussing an experiment employing electron counting for the measurement of the self-interference of individual electrons \cite{Gustavsson08a,Gustavsson08c}. These measurements may be seen as a solid-state implementation of the double-slit interference experiment \cite{Feynman06} conducted with individual electrons by Tonomura in 1989 \cite{Tonomura89}. The basic idea in these single-electron interference experiments is the appearance of the interference pattern as a result of building up the statistics of a large number of detection events.

\begin{figure}[tbp]
\includegraphics[width=8cm]{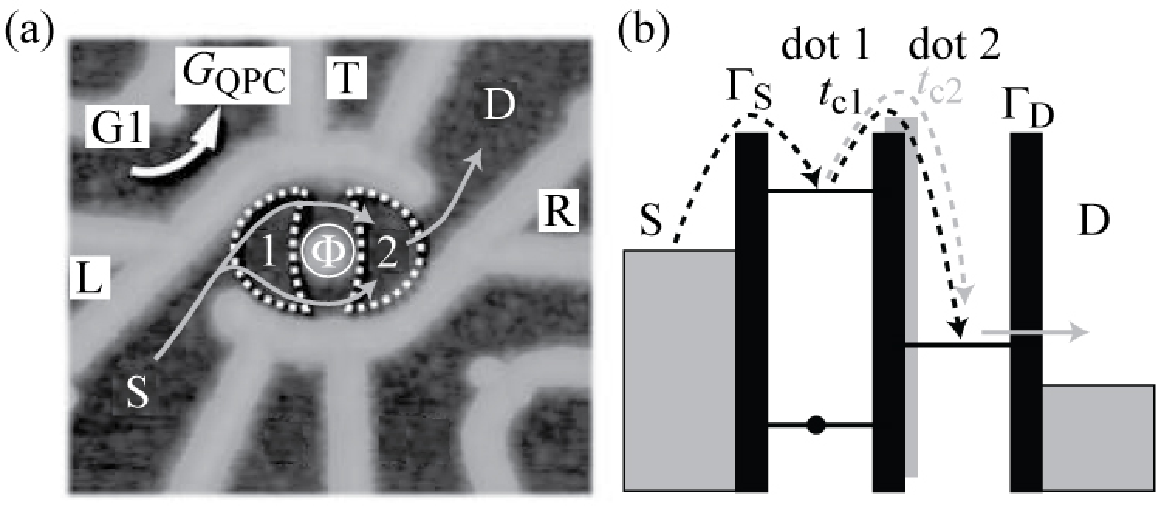}
\caption{(a) Scanning force microscope image of the double quantum dot with integrated charge read-out used for the self-interference experiment. (b) Schematic energy diagram of the double quantum dot system in the situation in which interference can be observed. The dashed arrows indicate the cotunneling-in process. Black and gray dashed arrows indicate the two alternative paths that enclose a magnetic flux $\Phi=BA$ used to tune their relative phase.}
\label{fig5}
\end{figure}

Figure\,\ref{fig5}(a) shows the sample used for this experiment. It is based on a shallow two-dimensional electron gas embedded in a Ga[Al]As heterostructure patterned by local anodic oxidation. The structure consists of two quantum dots as indicated by the dashed lines which are connected in series between a source (S) and drain (D) contact. In contrast to conventional double quantum dot systems there are two tunneling barriers connecting dot\,1 and 2, thereby allowing two spatially separate, parallel current paths that we denote the {\em upper} and {\em lower} path in the following. These two paths, together with the two dots, enclose an area through which a magnetic flux can be threaded in the experiment by applying an external magnetic field normal to the plane of the electron gas. The quantum point contact charge detector is capacitively coupled to the double quantum dot system, and its conductance $G_\mathrm{QPC}$ is read out via its own independent circuit. In-plane gates L, R, T, allow to tune the double dot system, and the gate G1 is used to set the operating point of the charge detector quantum point contact.

Using these gates, the system is carefully tuned into the state indicated schematically by the energy diagram in Fig.\,\ref{fig5}(b). The strengths of the tunneling coupling to source and drain, $\Gamma_\mathrm{S}$ and $\Gamma_\mathrm{D}$, were determined from measurements of the thermal noise and tuned to be below 15\,kHz, well below the bandwidth of the detector circuit. The tunneling coupling $\Gamma_\mathrm{c}\propto |t_\mathrm{c1}+t_\mathrm{c2}|^2$ between the two dots was measured to be a few gigahertz, i.e., beyond the time-resolution of the detector circuit. A finite source--drain bias voltage $V_\mathrm{DQD}$ applied to the double quantum dot system ensured $eV_\mathrm{DQD}\gg k_\mathrm{B}T$. Quantum dot 1 is tuned to a nonresonant situation in which an electron can traverse only by second order cotunneling processes from the source contact into quantum dot 2.
In this dot the electron stays for a sufficiently long time to be detected. Then it will leave dot 2 into the drain contact.

It is important to realize that the amplitude for the cotunneling process from the source contact into dot 2 is the sum of two amplitudes of the spatially separate upper and lower paths. In the presence of a magnetic field $B$ within the area $A$ enclosed by the two paths and the two quantum dots, the relative phase of these two amplitudes can be tuned, such that the cotunneling-in  rate $\Gamma_\mathrm{in}$ has the oscillatory Aharonov--Bohm contribution \cite{Aharonov59}
$ \Gamma_\mathrm{in} \propto \cos(eBA/\hbar) $. The most crucial condition for the interference experiment to work is that the tunneling coupling $t_\mathrm{c1}\approx t_\mathrm{c2}$, a condition which turned out to be hard to achieve experimentally. The tunneling-in rate $\Gamma_\mathrm{in}$ can be determined from detector time traces like that shown in Fig.\,\ref{fig2}(a) on the basis of single-electron tunneling events. In contrast, the tunneling-out rate $\Gamma_\mathrm{out}=\Gamma_\mathrm{D}$ which can similarly be determined is independent of $B$.

\begin{figure}[tbp]
\includegraphics[width=8cm]{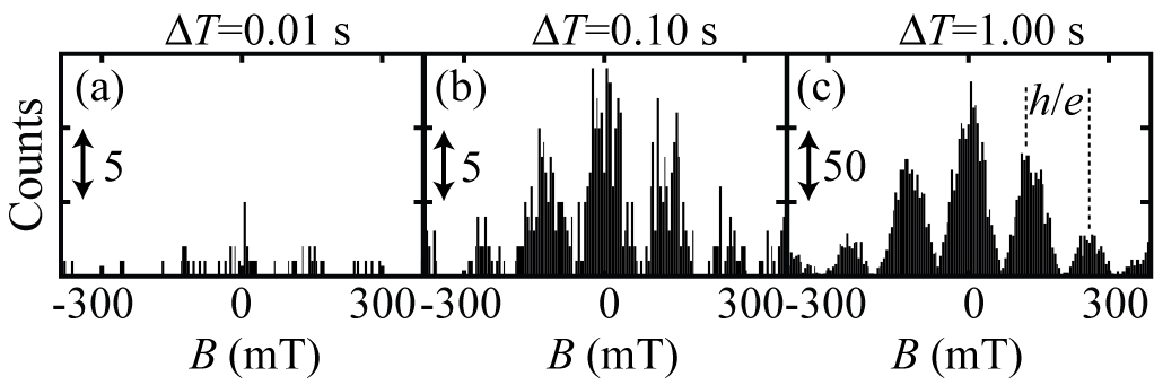}
\caption{Number of electrons that have traversed the double quantum dot interferometer (counts) measured at different magnetic fields $B$. Time traces of different lengths $\Delta T$ were used as the basis for the plotted histograms in (a), (b), and (c). In (c) it is indicated that adjacent interference maxima have a spacing one flux quantum $h/e$ per area $A$ enclosed by the interfering paths.}
\label{fig6}
\end{figure}

With the oscillatory modulation of $\Gamma_\mathrm{in}$ with magnetic field, the full counting statistics $P_N(\Delta T)$ becomes an oscillatory function of magnetic field. This implies that the counting experiment contains also the Aharonov--Bohm effect in the shot noise of the current through the quantum dot. In Fig.\,\ref{fig6} we show the number of electrons traversing the double dot system (counts) within a given time span $\Delta T$ at different magnetic fields $B$.
It can be seen in Fig.\,\ref{fig6}(a) that after short times, only a random pattern of counts is visible. Waiting ten times longer, in (b) the interference pattern can be seen to be still masked by significant statistical fluctuations. After a time of $\Delta T=1$\,s, the interference pattern is fully developed and statistical fluctuations are relatively weak.

Comparing our experiment with the interference experiment of Tonomura and coworkers \cite{Tonomura89}, there are important differences. In our experiment, the electrons are guided by the potential landscape created by sample fabrication along predefined paths, whereas Tonomura used an open geometry where the electrons were only slightly deflected by a biprism. Furthermore, in our experiment the observation of the electron's arrival cannot be made with a position-sensitive electron counting system. Here we rather use the magnetic field via the Aharonov--Bohm effect \cite{Aharonov59} to change the relative phase of paths that are fixed in space, and detect the arriving electrons at a fixed location (dot 2). As a consequence of the generalized Onsager symmetry relations in mesoscopic transport \cite{Buttiker86} the observed Aharonov--Bohm oscillations are necessarily even in magnetic field. In contrast to the original thought experiment by Aharonov and Bohm, but similar to previous Aharonov--Bohm interference experiments in metallic rings \cite{Webb85} and semiconductor nanostructures \cite{Timp87,Fuhrer01}, the magnetic field is not excluded from the spatial region of the electron's paths. However, Lorentz force effects have no significant influence, as long as the classical electronic cyclotron diameter is large compared to the area enclosed by the flux.

The visibility of the oscillations observed in Fig.\,\ref{fig6}(c) is close to 100\%. This implies that indeed $t_\mathrm{c1}=t_\mathrm{c2}$ is met very well in the experiment, but also decoherence effects are not significant.
We argue that this is due to the fact that the detector is not sensitive to the path that the electron took on the time scale of the fast cotunneling process \cite{Gustavsson08a}. This is in contrast to controlled dephasing experiments that use a charge detector to obtain `which path' information \cite{Buks98,Sprinzak00,Kalish04,Neder07}.

Finally we remark, that the sample used for this study of electron interference was used in the same regime to investigate cotunneling processes by counting in detail \cite{Gustavsson08}. It was possible to demonstrate the experimental equivalence of cotunneling and sequential tunneling into molecular states. In addition, the shot noise in the cotunneling current could be resolved. 

\section{Back-action}

It is a property of all quantum measurements that the detector unavoidably disturbs the measured system, a phenomenon called detector back-action. An equivalent way of looking at the same thing is to exchange the role of the measured system and the detector. We therefore ask the question, in which way the quantum dot (or double quantum dot) can detect electrons traversing the quantum point contact. It has been established in previous experiments that the shot noise of the quantum point contact current can couple into a system with coherent dynamics and act as a source for decoherence \cite{Buks98,Sprinzak00,Neder07}.

\begin{figure}[tbp]
\includegraphics[width=8cm]{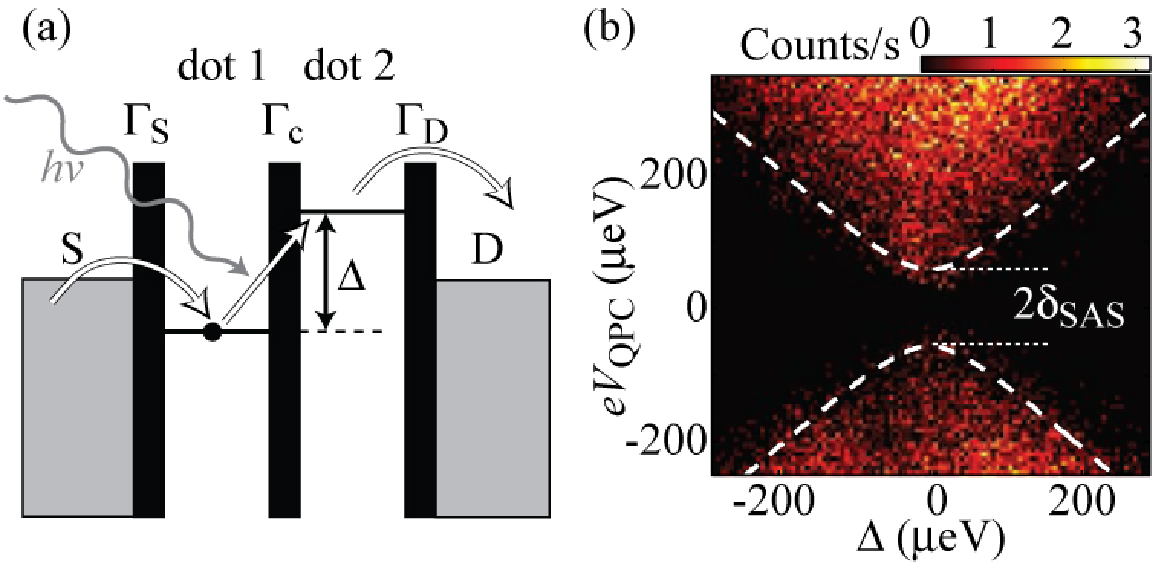}
\caption{(a) Schematic energy diagram showing the state of the double quantum dot in which it can be used to detect incident energy quanta with spectral resolution. (b) Number of  electrons leaving the double dot system per second measured as a function of $\Delta$ and the quantum point contact source--drain voltage $V_\mathrm{QPC}$.}
\label{fig7}
\end{figure}

We have performed complementary experiments using double quantum dot devices as a frequency selective detector to observe the shot noise in the quantum point contact \cite{Gustavsson07}. The principle of the double dot detector is shown in Fig.\,\ref{fig7}(a). No source--drain bias voltage is applied to the quantum dot. As a consequence, there is no net electron flow between the two contacts in thermodynamic equilibrium. The double dot may contain $N$ electrons in dot 1 and $M$ electrons in dot 2. The levels of the two dots are aligned in such a way that the $(N,M)$ charge state is the ground state of the system. The $(N-1,M+1)$ excited state belonging to the same total electron number in the double dot system is separated by an energy $\Delta$ which is larger than the thermal energy $k_\mathrm{B}T$. In this situation the double quantum dot is susceptible to energy quanta such as photons with an energy $h\nu=\Delta$.
Absorbing such an energy quantum excites the system into the state $(N-1,M+1)$, from which there are two decay routes: on one hand, an energy quantum with energy $h\nu=\Delta$ can be reemitted bringing the system back into the ground state. On the other hand, the system can decay into the state $(N-1,M)$, if one electron tunnels from dot 2 into the drain lead [see Fig.\,\ref{fig7}(a)]. In this case, another tunneling event will most likely follow, where an electron enters dot 1 from the source contact. This sequence of processes corresponds to the net transfer of a single electron from source to drain by virtue of the absorption of an energy quantum. This means that energy quanta absorbed by the quantum dot in this configuration can drive a net current at zero applied source--drain bias voltage. It turns out that energy quanta from a thermal bath, such as a photon bath, or the phonon bath of the host crystal, cannot drive this process, if these baths are in thermodynamic equilibrium with the electronic system in the source- and drain leads. However, it has been shown that this process can be driven, if the phonon bath temperature exceeds the temperature of the electron system in the contacts \cite{Gasser09}, or if nonequilibrium photons originating from the quantum point contact shot noise impinge onto the double quantum dot system \cite{Gustavsson07}.
If the energy separation $\Delta$ of the two involved states is changed, the frequency of the absorbed quanta can be changed. In principle, this allows the measurement of the spectral density of the incident energy quanta.

The particular experiments that were performed along these lines differ in principle. The experiment of Ref.\,\cite{Gustavsson07} uses the same quantum point contact detector that creates the shot noise to detect the tunneling of electrons from the double dot system into the leads in a time-resolved fashion. This detection method does not measure the induced double quantum dot current, but merely monitors charge leaving any of the dots into any of the two contacts. The result of this experiment is shown in Fig.\,\ref{fig7}(b). We first consider a cross section through the data along constant $V_\mathrm{QPC}\approx 300\,\mu$V. The count rate is maximum at $\Delta = 0$ and decays, if $|\Delta|$ is made larger. Along a cross section at constant $\Delta\approx 100\,\mu$eV, we notice that counts are only observed, if $|eV_\mathrm{QPC}|>\Delta$. This is compatible with the idea that the maximum energy a single electron can dissipate when traversing the quantum point contact is $|eV_\mathrm{QPC}|$. Only if this energy exceeds $\Delta$, the double quantum dot can be excited. The dashed line following the onset of the signal therefore gives the avoided crossing of the two involved quantum dot energy levels brought about by the finite tunneling coupling $\Gamma_\mathrm{c}$ of these states. The smallest separation at $\Delta=0$ is twice the symmetric--antisymmetric energy splitting $\delta_\mathrm{SAS}$. A detailed analysis of the data \cite{Gustavsson07} shows good agreement with a circuit model by Aguado and Kouwenhoven \cite{Aguado00} that couples quantum point contact shot noise to the double quantum dot system. Later experiments performed on a single quantum dot defined in an InAs nanowire strongly coupled to a quantum point contact in an underlying Ga[Al]As heterstructure two-dimensional electron gas, confirmed the model of shot noise coupling \cite{Gustavsson08b}.

\begin{figure}[tbp]
\includegraphics[width=8cm]{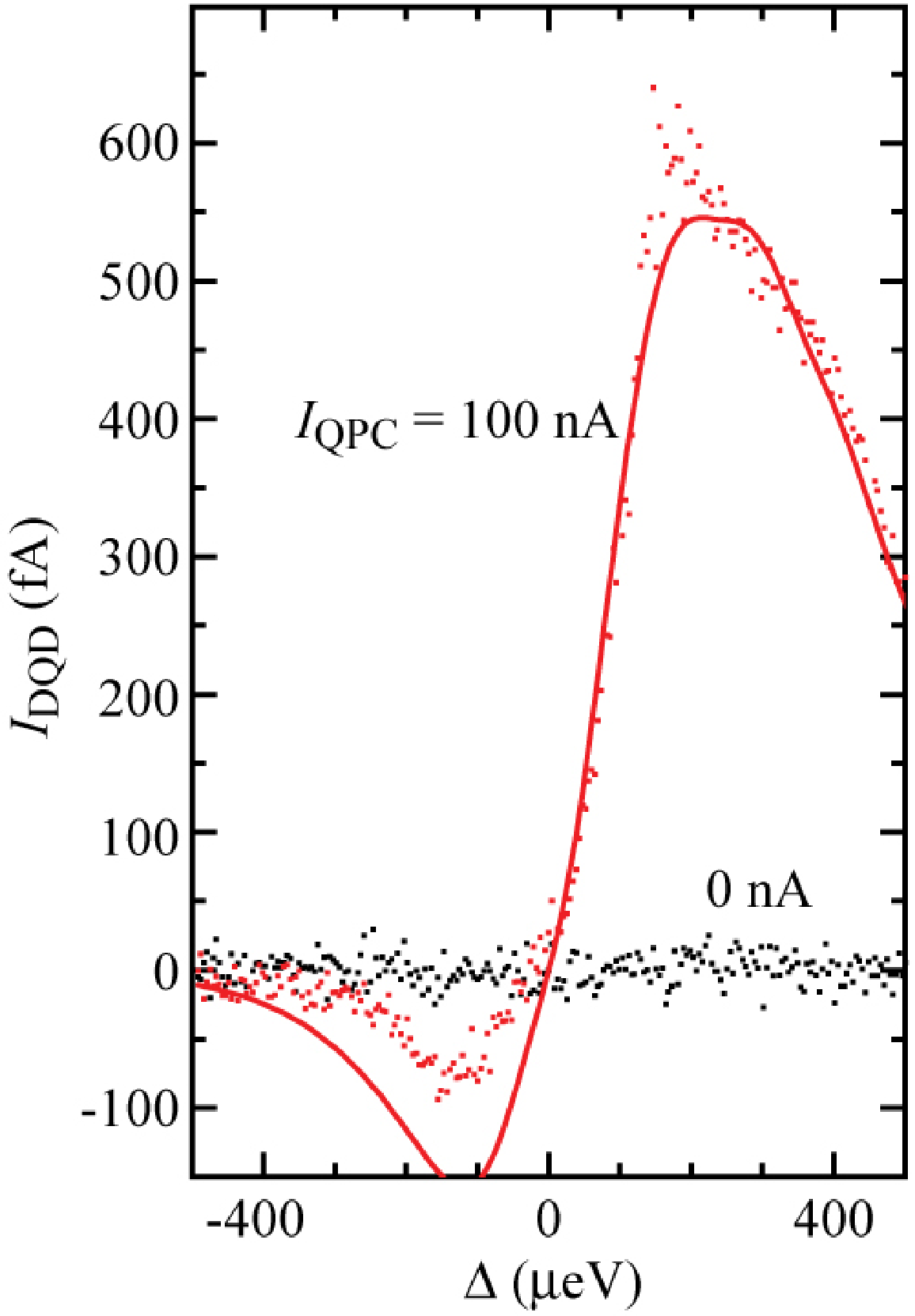}
\caption{Current driven through a double quantum dot at zero source--drain bias voltage as a result of heating the phonon bath with a quantum point contact. The solid curve is the result of a model calculation with the phonon bath temperature being the only fitting parameter.}
\label{fig8}
\end{figure}

The experiment in Ref.\,\cite{Gasser09}, however,  measures the time averaged double quantum dot current directly. In this experiment, the quantum point contact charge detector was only weakly Coulomb-coupled to the double quantum dot such that time-resolved charge detection was not possible. Furthermore the quantum point contact was operated on the first conductance plateau at $G_\mathrm{QPC}=2e^2/h$, where shot noise is suppressed. Therefore, the current shown in Fig.\,\ref{fig8} driven through the double quantum dot device at zero applied source--drain bias voltage cannot be explained by coupling to the shot noise of the quantum point contact. However, the experiment makes clear that the observed current is directly related to the strength of the quantum point contact current $I_\mathrm{QPC}$. Similar experiments have been conducted in Ref.\cite{Khrapai06} and interpreted in terms of a double dot quantum ratchet driven by the quantum point contact.

In contrast to the experiment described before, where a maximum number of counts was detected at $\Delta=0$, the measured $I_\mathrm{DQD}$ is zero at $\Delta = 0$ and shows a pronounced maximum at finite detuning $\Delta$. If $\Delta$ is very small, the avoided crossing of the two energy levels brought about by the finite coupling $\Gamma_\mathrm{c}$ between the two dots leads to a situation where electrons tend to tunnel with similar probability into the source and into the drain contact. This situation results in zero current. In contrast, large positive $\Delta$ leads to a current flowing preferentially from source to drain, whereas large negative $\Delta$ leads to a current in the opposite direction explaining the observed sign change of $I_\mathrm{DQD}$ with $\Delta$.

 It turns out that the data shown in Fig.\,\ref{fig8} is in accordance with a model \cite{Gasser09} that takes coupling of the double quantum dot to a phonon bath into account. For simplicity, the occupation of states in the phonon bath is assumed to be described by the equilibrium Bose--Einstein distribution function at a temperature $T_\mathrm{ph}$. However, it is implicitly assumed that $T_\mathrm{ph}$ can be increased by driving a higher current through the quantum point contact. The double quantum dot absorbs phonons from the bath at the particular energy $\Delta$. The high energy cut-off of $I_\mathrm{DQD}$ seen in Fig.\,\ref{fig8} is the result of the decreasing phonon occupation at higher energies. The solid line in Fig.\,\ref{fig8} shows the result of the model calculation where $T_\mathrm{ph}$ was taken as the fitting parameter.

\section{Concluding remarks}

Within the past five years the use of quantum point contact charge detectors in research related to semiconductor nanostructures has seen an unprecedented rise in popularity. Experiments with these
detectors have given completely new insights into the physics of quantum dots, and they have allowed to
access parameter regimes that were inaccessible before. They can be easily integrated on chip, are
weakly invasive, and allow to explore even the spin degree of freedom via spin-to-charge conversion. Currently quantum point contact charge detectors are state-of-the-art devices for qubit research. On the basis of these
developments we are confident to anticipate the development of further fascinating experiments leading to intriguing results and enlightening insights into the world of quantum dots in particular, and nanostructures in general.

\appendix

\section*{Acknowledgements}
We thank W. Wegscheider, M. Reinwald, D.C. Driscoll, and A.C. Gossard for supplying the world class shallow two-dimensional electron gases that were needed for fabricating the samples discussed in this paper.
Financial support by the Swiss National Science Foundation is gratefully acknowledged.

\end{document}